\documentclass{IET/cta-author}

\usepackage{amsmath,amssymb}
\usepackage{newtxtext,newtxmath}
{}
{}
{}
\usepackage{algorithmic}
\usepackage{graphicx}
\usepackage{textcomp}
\usepackage{array}
\usepackage{amsmath,amssymb,amsfonts}
\usepackage{algorithmic}
\usepackage{graphicx}
\usepackage{textcomp}
\usepackage{multirow}
\usepackage{array}

\newcolumntype{P}[1]{>{\centering\arraybackslash}p{#1}}
\usepackage{booktabs}
\usepackage[T1]{fontenc}
\usepackage[utf8]{inputenc}

\begin{document}

\title{Dynamic Attitude Estimation Improvement for Low-cost MEMS IMU by Integrating Low-cost GPS}

\author{\au{Guiqiu Liao$^{2,3,1}$}
\au{Jiankang Zhao$^1$}
\au{Chao Cui$^1$}
\au{Haihui Long$^1$}
\au{Jianbin Zhu$^4$}
\au{Achraf Djerida$^1$}
}

\address{\add{1}{School of Electronic Information and Electrical Engineering, 
Shanghai Jiao Tong University, Shanghai, China
}
\add{2}{Imaging, Robotics, Remote Sensing \& Health Department, ICube, 
The University of Strasbourg, Strasbourg, France
}
\add{3}{ALTAIR Robotics Laboratory, Department of Computer Science,
 The University of Verona, Verona, Italy
}
\add{4}{Shenzhen High Speed Engineering Consulting Co., Ltd., Shenzhen, China}
\email{guiqiu.liao@icube.unistra.fr}}

\begin{abstract}
This paper proposes a low-cost six Degree-of-Freedom (6-DOF) navigation system for small aerial robots based on the integration of Global Position System (GPS) receiver with sensors of inertional Microelectromechanical Systems (MEMS). In the problem of fusing Inertial Measurement Unit (IMU) with low-cost GPS, the effect of time synchronization error on attitude estimation is concerned. A fusion algorithm which can estimate the motion states and the time synchronization error simultaneously is proposed. This algorithm adds a time estimation loop to improve estimation accuracy. Compared with another states augmented estimation approach, this method has the advantages of lower computation burden, avoidance of the discretization error in the low sample rate. The estimation algorithm is implemented in an low-cost embedded microprocessor where the update rate of algorithm can achieve more than 100 Hz, and therefore high-performance computational units are not necessary. In robotic experiment, the proposed algorithm serves as the navigation solution for a small aerial robot. The accuracy and reliability of the self-designed system are tested when the robot is moving with significant acceleration.
\end{abstract}

\maketitle
\section{Introduction}
Attitude angle  estimation is vital for robotic applications such as automatic driving and aerial machines. Recently, low-cost GPS/IMU integrated navigation systems have become a popular method to provide navigation solution for motion controlling. At present, the output position error of low-cost GPS receivers is about 2 meters, and the accuracy of speeds can reach 5cm/s \cite{b1}. Although the navigation errors of GPS do not accumulate over time, its anti-interference ability is poor, any shelter can affect its output. On the other hand, Attitude and Heading Reference System (IMU) integrating MEMS gyroscope, accelerometer and magnetometer has a good autonomy and real-time performance. A strap-down IMU can provide angle rates, specific forces and heading angle measurements for a moving platform at high rates, typically 400 times per second \cite{b2}. The small size of low-cost IMU makes it indispensable for small unmanned aerial robots’ navigation. However, the inertial sensors’ drifts and biases, in the IMU, will accumulate over time. Furthermore, when the acceleration of a vehicle is changing frequently, the rotation attitude estimated by IMU is not as accurate as the attitude estimated when the platform is still, therefore the acceleration calculation errors will be amplified in this situation \cite{b3}. Since the GPS has been developed to be a global universal navigation system, it is commonly used for integration with IMU \cite{b4}. The GPS/IMU integration can improve the accuracy and instantaneity of motion estimation, and it removes the effect of specific force measurements, which resulted from linear acceleration in the attitude estimation. However, the time synchronization error between GPS and IMU is a predominant factor affecting the fusion algorithm. To ensure the performance of GPS/IMU navigation, it is crucial to take the problem of time synchronization into consideration. 

In a GPS/IMU loosely coupled system, the GPS update rate is relatively low (5~20Hz), and the update rate of IMU is relatively high. In addition, for the GPS navigation, after receiving the signal from the satellites, it has to go through the frequency conversion and position solution \cite{b5}. Hence the GPS receiver output data has poor instantaneity compared to IMU measurements. When GPS and IMU output data at the same instant, position and velocity provided by GPS receiver lag behind the IMU data.



Research on the impact of synchronization error in integrated systems started early \cite{b6}. Some researchers have tried to use one pulse-per-second (1PPS) signal detection method to solve the time synchronization problem of the GPS/Strap-down Inertial Navigation System (SINS) \cite{b7}.Some time synchronization devices by hardware approaches can be seen in  Skaloud \cite{b8,b9,b10,b11}, which could well calculate the time synchronization errors caused by different data update rates and transmission efficiencies. However, this method cannot guarantee that the aligned data represents the motion of the platform at the same instant. The software approach has been tried and proved to be a feasible method. In the research of Bouvet \cite{b12}, high precision Real-Time Kinematic difference Global Positioning System (RTK-GPS) was put forward to estimate the GPS delay based on the position error model, where the estimated result is roughly 0.8s~1s. The stability of delayed systems is further investigated in \cite{b13} and \cite{b14}.

Recently, researches on time-delay in nonlinear fusion systems that utilize intelligent fusion techniques and Kalman filter have increased. In \cite{b15} a small-gain approach was used to realize a class of global exponential stable nonlinear observers with sampled and delayed measurements, robust towards measurement errors and sampling schedule perturbations. In \cite{b16} a class of nonlinear predictors for delayed measurements with a known and constant delay is proposed. The nonlinear observer consists of several couples of filters estimating the state vector at some delayed time instant differing from the previous by a small fraction of the overall delay. In \cite{b17} research work, an observer for delayed GPS aided INS navigation is based on USGES nonlinear estimator \cite{b18}, and it is implemented to estimate the Position, Velocity and Attitude(PVA) for fixed-wing unmanned aerial vehicle (UAV) . \cite{b19} proposed a modified filtering framework for randomly delayed measurements, which is a generalized approach for arbitrary time-step delayed measurements. For some intelligent nonlinear fusion estimators, the possibly varying delay of GPS measurements is also an important parameter. \cite{b20} concerned multi-sensor optimal $H_{\infty}$ fusion filter for a class of nonlinear intelligent systems with time delays. A unified model consisting of a linear dynamic system and a bounded static nonlinear operator is employed to describe these systems, such as neural networks and Takagi and Sugeno (T-S) fuzzy models. \cite{b21} applys autoregressive neural network fusion architecture to fuse low-cost GPS and inertial measurement unit (IMU), the fusion architecture is a non-linear method that takes the variable delay between GPS measurement epochs into account.

Sensor fusion algorithms based on linear Kalman filter, Extended Kalman Filter (EKF) or Unscented Kalman Filter (UKF) are widely used in the navigation of mobile robots. Among these three filters, the linear Kalman filter requires the least computing resource, and the computational demand of UKF is highest \cite{b22}. \cite{b23,b24} study the joint state and parameter estimation problem for a linear state-space system with sensors time-delay based on Kalman filter. This method is designed for linear dynamic systems, and it is not suitable for the nonlinear PVA estimation navigation system with inertial sensors. \cite{b25,b26,b28} use the method of augmenting the states of EKF to estimate the time synchronization error in a loosely coupled system, where a second-order polynomial (or higher-order polynomial) is used to reflect the relationship between time synchronization error and other states. However, when the motion acceleration of the vehicle is changing dramatically, the error in the linearization and discretization will be amplified, and then it can affect the robustness of the filter. In addition, these methods are implemented in the situation where GPS update rates are relatively high, however, as for the low-cost off-the-shelf GPS receiver with low update rates (no more than 10 Hz), the accuracy and instantaneity requirements of the estimation may not be satisfied simultaneously, which is infeasible for dynamic control systems. In another recent related study, \cite{b27} modifys the UKF for arbitrary time delayed measurements. In this method, there is no consideration on the physical property of measurements and states, and it provides a more accurate estimation compared to the ordinary UKF in simulation with randomly delayed measurements. Due to the computational burden, this method based on UKF may be not practical for the navigation system with limited computation resources.

In this article, we develop a algorithm to fuse low-cost GPS and IMU sensor data to estimate rotorcraft’s 6-DOF motion.
The objective of this work is to improve the attitude estimation accuracy of IMU in dynamic circumstance by integrating an off-the-shelf GPS. 
Conventionally, modern GPS receivers provide to user special time pulses for hardware synchronization, however new software time synchronization methods are considered in the article.
The linear motion acceleration, which is computed based on different systems’ measurements, is the key to estimate the GPS delay. However, it will be a challenge to guarantee precision of the motion acceleration estimation using the accelerometers data of IMU, because the acceleration estimation is affected by the attitude estimation. In addition to that, the influence of the synchronization error will affect the attitude estimation accuracy in dynamic situation. 
To solve this problem, a feasible closed-loop filter with feedback is designed to estimate the time synchronization error, attitude and motion acceleration simultaneously.
We mainly compare the proposed algorithm with an conventional software approach - augmented EKF \cite{b28}, where the estimator states are augmented to estimate the time delay. Our new proposed approach takes the linearization and discretization errors, which resulted from the low sample rate of observations into account. To avoid these error we seperate the time estimation loop from the 6-DOF motion estimation, which guarantees the filter robustness at high dynamic conditions. Moreover, the separated loop has the ability to get rid of uncorrelated measurements. Therefore, the estimation instantaneity can be maintained running on a low-cost micro processor, even with a low sample rate GPS receiver.

This paper is organized as follows: In Section 2, the fusion algorithm for robot navigation based on extended Kalman filter is presented, and two solutions on time synchronization are presented and compared. In Section 3, experiments results of the proposed method and other algorithms are presented. Finally, results summaries are given and conclusions are drawn in Section 4.

\section{Methods}
We compare 2 algorithm architectures based on EKF on the angle estimation improvement with IMU and GPS. The first approach is a form of Augment EKF for navigation states estimation, which adds the time lag error in the states vector. Another approach is to estimate the time delay in another independent loop.

\subsection{Dynamic States Estimation with EKF}
For different sensor fusion algorithms, the effect of time synchronization errors will be varied. In most estimators, based on the estimation on multiple states influences each other mutually. Taking this into consideration, this paper introduces a data fusion algorithm based on an extend Kalman filter. The 16 states that include the attitude, motion and sensor errors, are expressed as:
\begin{equation}
 \boldsymbol{x} = {[\begin{array}{*{20}{c}}
  \boldsymbol{q}&\boldsymbol{r}&\boldsymbol{v}&{{\boldsymbol{b}_f}}&{{\boldsymbol{b}_\omega }} 
\end{array}]^T}
\label{eq_state}\end{equation}
where $\boldsymbol{q} = {\left[ {\begin{array}{*{20}{c}}
  {{q_0}}&{{q_1}}&{{q_2}}&{{q_3}} 
\end{array}} \right]^T}$
  is the quaternion which expresses the rotation between the body coordinate frame and navigation frame,   and are the position and velocity in navigation frame.   and   are the bias errors of the accelerometer and gyroscope outputs.
  $\boldsymbol{r} = {\left[ {\begin{array}{*{20}{c}}
  {{r_n}}&{{r_e}}&{{r_d}} 
\end{array}} \right]^T}
$ and $\boldsymbol{v} = {\left[ {\begin{array}{*{20}{c}}
  {{v_n}}&{{v_e}}&{{v_d}} 
\end{array}} \right]^T}$ are the position and velocity in navigation frame.
$\boldsymbol{b}_f = {[\begin{array}{*{20}{c}}
  {{b_{fx}}}&{{b_{fy}}}&{{b_{fz}}} 
\end{array}]^T}
$ and ${\boldsymbol{b}_\omega } = {[\begin{array}{*{20}{c}}
  {{b_{\omega x}}}&{{b_{\omega y}}}&{{b_{\omega z}}} 
\end{array}]^T}$
are the bias errors of the accelerometer and gyroscope measurements.

The non-linear system model is described by the following kinematic relationships \cite{b_trans} :
\begin{equation}
\dot{ \boldsymbol{x} }= \left[ {\begin{array}{*{20}{c}}
  {{{\dot q}_0}} \\ 
  {{{\dot q}_1}} \\ 
  {{{\dot q}_2}} \\ 
  {{{\dot q}_3}} \\ 
  {{{\dot r}_n}} \\ 
  {{{\dot r}_e}} \\ 
  {{{\dot r}_d}} \\ 
  {{{\dot v}_n}} \\ 
  {{{\dot v}_e}} \\ 
  {{{\dot v}_d}} \\ 
  {{{\dot b}_{fx}}} \\ 
  {{{\dot b}_{fy}}} \\ 
  {{{\dot b}_{fz}}} \\ 
  {{{\dot b}_{\omega x}}} \\ 
  {{{\dot b}_{\omega y}}} \\ 
  {{{\dot b}_{\omega z}}} 
\end{array} } \right]{\text{ = }}\left[ {\begin{array}{*{20}{c}}
  {\frac{1}{2}\left[ {\begin{array}{*{20}{r}}
  { - {{\hat q}_1}}&{ - {{\hat q}_2}}&{ - {{\hat q}_3}} \\ 
  {{{\hat q}_0}}&{ - {{\hat q}_3}}&{{{\hat q}_2}} \\ 
  {{{\hat q}_3}}&{{{\hat q}_0}}&{ - {{\hat q}_1}} \\ 
  { - {{\hat q}_2}}&{{{\hat q}_1}}&{{{\hat q}_0}} 
\end{array}} \right]\left[ {\begin{array}{*{20}{c}}
  {\tilde \omega _x  - {{\hat b}_{\omega x}}} \\ 
  {\tilde \omega _y  - {{\hat b}_{\omega y}}} \\ 
  {\tilde \omega _z  - {{\hat b}_{\omega z}}} 
\end{array}} \right]} \\ 
  {\begin{array}{*{20}{c}}
  {{{\hat v}_n}} \\ 
  {{{\hat v}_e}} \\ 
  {{{\hat v}_d}} 
\end{array}} \\ 
  {{C}_n^b\left( {\left[ {\begin{array}{*{20}{c}}
  {\tilde f_x  - {{\hat b}_{fx}}} \\ 
  {\tilde f_y  - {{\hat b}_{fy}}} \\ 
  {\tilde f_z  - {{\hat b}_{fz}}} 
\end{array}} \right] + \left[ {\begin{array}{*{20}{c}}
  0 \\ 
  0 \\ 
  g 
\end{array}} \right]} \right)} \\ 
  {\begin{array}{*{20}{c}}
  0 \\ 
  0 \\ 
  0 \\ 
  0 \\ 
  0 \\ 
  0 
\end{array}} 
\end{array}} \right]
\label{eq_trans}\end{equation}
\begin{equation}
{C}_n^b = \left[ {\begin{array}{*{20}{c}}
  {1 - 2(\hat q_2^2 + \hat q_3^2)}&{2(\hat q_1^{}\hat q_2^{} - \hat q_3^{}\hat q_0^{})}&{2(\hat q_1^{}\hat q_3^{} + \hat q_2^{}\hat q_0^{})} \\ 
  {2(\hat q_1^{}\hat q_2^{} + \hat q_3^{}\hat q_0^{})}&{1 - 2(\hat q_1^2 + \hat q_3^2)}&{2(\hat q_2^{}\hat q_3^{} - \hat q_1^{}\hat q_0^{})} \\ 
  {2(\hat q_1^{}\hat q_3^{} + \hat q_2^{}\hat q_0^{})}&{2(\hat q_2^{}\hat q_3^{} + \hat q_1^{}\hat q_0^{})}&{1 - 2(\hat q_1^2 + \hat q_2^2)} 
\end{array}} \right]
\label{eq_C}\end{equation}
where the estimated rotation matrix ${C}_n^b$ is expressed by estimated quaternion $\tilde{\boldsymbol q}$. in equation (\ref{eq_trans}), where the differential of the rotation quaternion $\dot{\boldsymbol{q}}$ depends on the rotation rates rotation rates $\tilde{\boldsymbol{\omega}}$ measured by gyros and the estimated measurement bias $\hat{\boldsymbol{b}}_{\omega}$ (throughout this article, the over-bar denotes a predicted variable, the hat denotes an estimated variable, the dot denotes a differential value and the tilde denotes a measured variable); The velocity differential $\dot{\boldsymbol{v}}$ is calculated using the specific force $\tilde{\boldsymbol{f}} $ measured by accelerometers, the estimated specific force measurement bias $\hat{\boldsymbol{b}}_f$ and the estimated rotation matrix ${C}_n^b$, which is under the assumption that the transport rate between navigation coordinates and earth frame is negligible \cite{b_godha}. The inertial sensor biases are considered as constant values with slow drift, so that their differential $\dot{\boldsymbol b}_{\omega}$ and $\dot{\boldsymbol b}_{f}$ in prediction are considered as zero.

A form of discretized states prediction equation can be given as:
\begin{equation}
{\bar{\boldsymbol x}_k} =  {\hat{ \boldsymbol x}_{k - 1}} + {t_s} \cdot {{\dot { \boldsymbol x}}_{k - 1}} 
\label{eq_pre}\end{equation}
where $\bar{\boldsymbol x}_k$ represents the prediction of the navigation states, $\hat{ \boldsymbol x}_{k - 1}$ represents the estimated states of last iteration update. As the states prediction function relies on the inertial sensors data, the prediction interval time $t_s$ equals to the inertial sensors update interval time.
We calculate the Jacobian matrix $ {F}_{k-1}$ of $\dot{ \boldsymbol x}_{k - 1}$ \cite{b_kf} to get the linearized mapping, and then use the method from \cite{b_L} to discrete the $ {F}_{k-1}$ to obtain a state transition matrix $ {A}_{k-1}$. $ {A}_{k-1}$ is used for discrete noise covariance prediction of the Kalman filter:
\begin{equation}
{{\bar{P}}_{k}}={{A}_{k-1}}{{\hat{P}}_{k-1}}{{A}_{k-1}}+{{Q}_{k-1}}
\label{eq_Pp}\end{equation}
where $\bar{P}_{k}$ is the prediction of current states co-variance matrix, $\hat{P}_{k-1}$ is the estimated states co-variance matrix for the last algorithm cycle, ${Q}_{k-1}$ is the desecrated co-variance matrix of process noise.

As for the observational measurements $\tilde{\boldsymbol z }_{k}={{[\begin{matrix}
   \tilde{\boldsymbol m}_{k}^{b} & \tilde{\boldsymbol r}_{k}^{n} & \tilde{ \boldsymbol v}_{k}^{n}  \\
\end{matrix}]}^{T}}$
 which are used for correction of EKF prediction, is composed by the magnetometer’s measurements $\tilde{\boldsymbol z }_{k}$, the position measurements $\tilde{\boldsymbol r}_{k}^{n}$ and the velocity measurements $\tilde{ \boldsymbol v}_{k}^{n}$ of GPS receiver. Instead of using the magnetometer output directly,  We calculate it based on attitude angle of last estimation step:
\begin{equation}
\tilde{\boldsymbol m}_{k}^{b}=  {{{\hat{C}}}_{\phi }}\cdot {{{\hat{C}}}_{\theta }}\cdot {{{\tilde{C}}}_{\psi }}\cdot {{\boldsymbol m}^{n}} 
\end{equation} 
\begin{equation}
{{\hat{C}}_{\phi }}=  \left[ \begin{matrix}
   1 & 0 & 0  \\
   0 & \cos \hat{\phi } & \sin \hat{\phi }  \\
   0 & -\sin \hat{\phi } & \cos \phi   \\
\end{matrix} \right]
\end{equation} 
\begin{equation}
{{\hat{C}}_{\theta }}= \left[ \begin{matrix}
   \cos \hat{\theta } & 0 & -\sin \hat{\theta }  \\
   0 & 1 & 0  \\
   \sin \hat{\theta } & 0 & \cos \hat{\theta }  \\
\end{matrix} \right]
\end{equation}
\begin{equation}
{{\tilde{C}}_{\psi }}=\left[ \begin{matrix}
   \cos {{{\tilde{\psi }}}_{mag}} & \sin {{{\tilde{\psi }}}_{mag}} & 0  \\
   -\sin {{{\tilde{\psi }}}_{mag}} & \cos {{{\tilde{\psi }}}_{mag}} & 0  \\
   0 & 0 & 0  \\
\end{matrix} \right]
\end{equation}
where the rotation matrices ${{\hat{C}}_{\phi }} $ and  ${{\hat{C}}_{\theta }}$ are generated by the estimated roll and pitch angles  $\hat{\phi } $ and  $\hat{\theta } $ calculated by quaternion  $\hat{q} $; the third rotation matrices  ${{\tilde{C}}_{\psi }}$ is generated by the magnet heading angle  ${{\tilde{\psi }}_{mag}}$; The normalized local earth magnet field in NED frame  ${{\boldsymbol m}^{n}} $ is calculated with magnet declination angle  ${{\theta }_{d}}$ and inclination angle ${{\theta }_{i}}$: ${{\boldsymbol m}^{n}}={{[\begin{matrix}
   \sin {{\theta }_{i}}\cos {{\theta }_{d}} & \sin {{\theta }_{i}}\sin {{\theta }_{d}} & \cos {{\theta }_{i}}  \\
\end{matrix}]}^{T}}$
, which reflect the deviation between true north and magnetic field direction. For example, the magnetic declination is $\text{-}{{6}^{\circ }}{2}'$ and the inclination is ${{47}^{\circ }}1{4}'$ in Shanghai, China \cite{b_wmm}. In this way, the magnetometer data will only correct the estimation of the heading angle, without affecting the attitude estimation.

We first model the relationship between the motion states and the observation assume that the GPS is aligned with the IMU:
\begin{equation}
{{\tilde{\boldsymbol z}}_{{}}}=\left[ \begin{matrix}
   \tilde{m}_{x}^{{}}  \\
   \tilde{m}_{y}^{{}}  \\
   \tilde{m}_{z}^{{}}  \\
   \tilde{r}_{n}^{{}}  \\
   \tilde{r}_{e}^{{}}  \\
   \tilde{r}_{d}^{{}}  \\
   \tilde{v}_{n}^{{}}  \\
   \tilde{v}_{e}^{{}}  \\
   \tilde{v}_{d}^{{}}  \\
\end{matrix} \right]=h({{\boldsymbol x}_{k}})+{{\boldsymbol e}_{k}}=\left[ \begin{matrix}
   C{{_{n}^{b}}^{T}}{{\boldsymbol m}^{n}}  \\
   r_{n}^{{}}  \\
   r_{e}^{{}}  \\
   r_{d}^{{}}  \\
   v_{n}^{{}}  \\
   v_{e}^{{}}  \\
   v_{d}^{{}}  \\
\end{matrix} \right]+{{ \boldsymbol e}_{k}}
\end{equation}

By calculating the Jacobian matrix of  $h({{\boldsymbol x}_{k}}) $with respect to the $16\times 1 $states vector  ${{\boldsymbol x}_{k}} $, the linearized  $9\times 16$ observation matrix  ${{H}_{k}} $ is expressed as:
\begin{equation}
{{H}_{k}}=\left[ \begin{matrix}
H_{3\times 4}^{m} & {{0}_{3\times 3}} & {{0}_{3\times 3}} & {{0}_{3\times 6}}  \\
{{0}_{3\times 4}} & {{I}_{3\times 3}} & {{0}_{3\times 3}} & {{0}_{3\times 6}}  \\
{{0}_{3\times 4}} & {{0}_{3\times 3}} & {{I}_{3\times 3}} & {{0}_{3\times 6}}  \\
\end{matrix} \right]
\end{equation}
where  $H_{3\times 4}^{m} $ represents the  $3\times 4 $ Jacobian matrix of the magnetic vector in the body frame with respect to the rotation quaternion  $\boldsymbol q $.  ${{I}_{3\times 3}} $ represents the  $3\times 3 $ identity matrix. So the EKF gain  ${{K}_{k}} $ is updated by:
\begin{equation}
{{K}_{k}}=({{P}_{k}}H_{k}^{T}){{(H_{k}^{T}{{P}_{k}}H_{k}^{T}+{{R}_{k}})}^{-1}}
\label{eq_K}\end{equation}
where ${{R}_{k}}$ is the measurement covariance matrix. When new data comes, the states ${\bar{\boldsymbol x}_{k}} $ predicted by inertial sensors can be corrected by the new measurements observation vector as:
\begin{equation}
{{\hat{\boldsymbol x}}_{k}}= {{\bar{\boldsymbol x}}_{k}}+{{K}_{k}}\left( {{{\tilde{\boldsymbol z}}}_{k}}-h({{{\bar{\boldsymbol x}}}_{k}}) \right)
\label{eq_crct}\end{equation}
where ${{\hat{\boldsymbol x}}_{k}}$ is the final states estimation including attitude angle, linear velocity, position and measurement biases.

\subsection{Time Delay compensation}

\subsubsection{Augmented EKF}
In real situation when GPS is fused with IMU, when GPS has significant delay, the measurement could be modeled based upon a send order Taylor expansion of position and velocity around each sampling time:
\begin{equation}
\tilde{\boldsymbol z}^d =\left[ \begin{matrix}
\tilde{m}_{x}^{{d}}  \\
\tilde{m}_{y}^{{d}}  \\
\tilde{m}_{z}^{{d}}  \\
\tilde{r}_{n}^{{d}}  \\
\tilde{r}_{e}^{{d}}  \\
\tilde{r}_{d}^{{d}}  \\
\tilde{v}_{n}^{{d}}  \\
\tilde{v}_{e}^{{d}}  \\
\tilde{v}_{d}^{{d}}  \\
\end{matrix} \right]=h^d({{\boldsymbol x}_{k}})+{{\boldsymbol e}_{k}}=\left[ \begin{matrix}
C{{_{n}^{b}}^{T}}{{\boldsymbol m}^{n}}  \\
  {\boldsymbol  r}_{k} -{\boldsymbol v}_{k}{{\tau }_{f}}+ {\boldsymbol a}_{k} {{\tau }_{f}}^{2}/2  \\
   {\boldsymbol v}_{k}-{\boldsymbol a}_{k}{{\tau }_{f}}   \\
\end{matrix} \right]+{{ \boldsymbol e}_{k}}
\end{equation}
where ${{\tau }_{f}}$is the time synchronization error between GPS output and raw IMU sensor data, $\boldsymbol r_k$ and $\boldsymbol v_k$ are the position and velocity vectors respectively in the navigation frame.

To estimate the ${{\tau }_{f}}$, one way is to augment the 16 EKF states vector $\boldsymbol x$ to a 17 states vector which contains the delay:
\begin{equation}
{{\boldsymbol x}^{d}}=\left[ \begin{matrix}
   \boldsymbol x  \\
   {{\tau }_{f}}  \\
\end{matrix} \right]
\end{equation}

Correspondingly, by calculating new Jacobian matrix of  $h^d({{\boldsymbol x}_{k}}) $ with respect to the $17\times 1 $states vector  ${\boldsymbol x}_k^d$, a new linearized  $9\times 17$ observation matrix $H_k^d$ considering the delay error is obtained. Then in the EKF $H_k^d$ will be applied in equation (\ref{eq_K}) and (\ref{eq_crct}) to correct the navigation states prediction.

\subsubsection{Separated Time Estimation Loop}
\begin{figure*}[!ht]
\centerline{\includegraphics[width=0.8\textwidth]{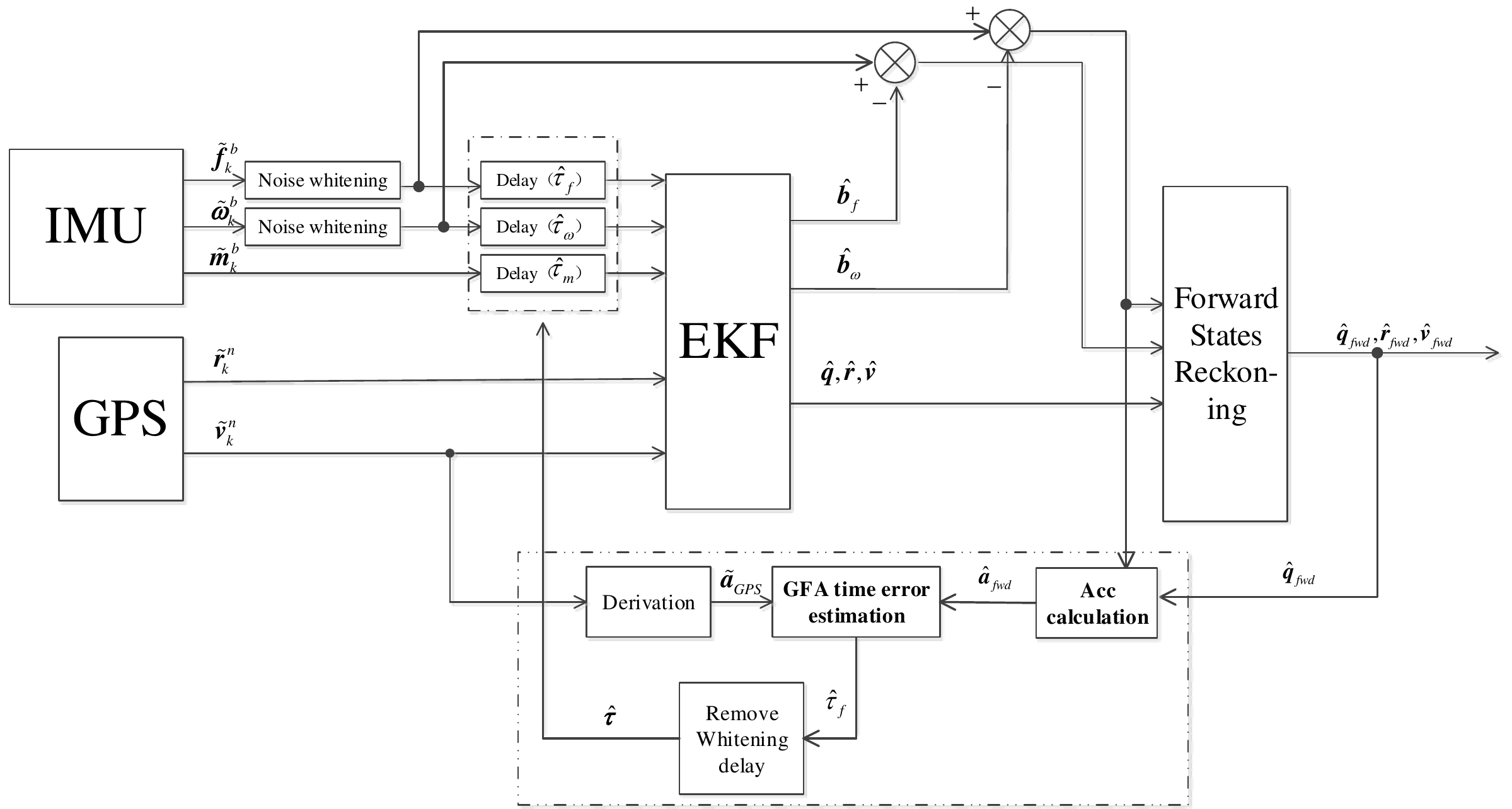}}
\caption{Block diagram of the proposed GPS/IMU fusion algorithm.}
\label{fig_method}
\end{figure*}

\begin{figure}[!ht]
\centerline{\includegraphics[width=1.0\columnwidth]{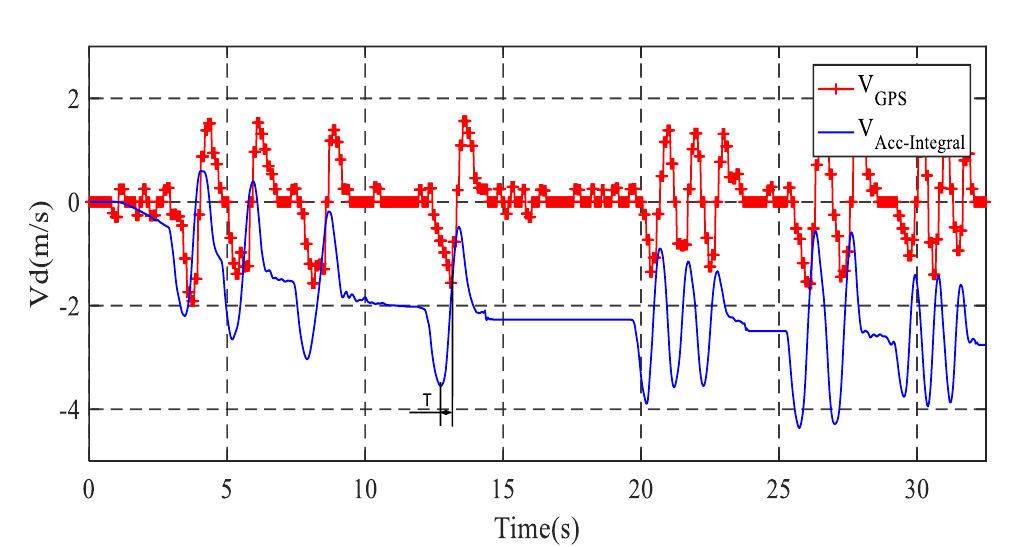}}
\caption{Down velocities in NED navigation coordinates from Ublox-M8 GPS
 and calculation with MPU9250 IMU sensor data
}
\label{fig_lag}
\end{figure}

\begin{figure}[!ht]
\centerline{\includegraphics[width=1.0\columnwidth]{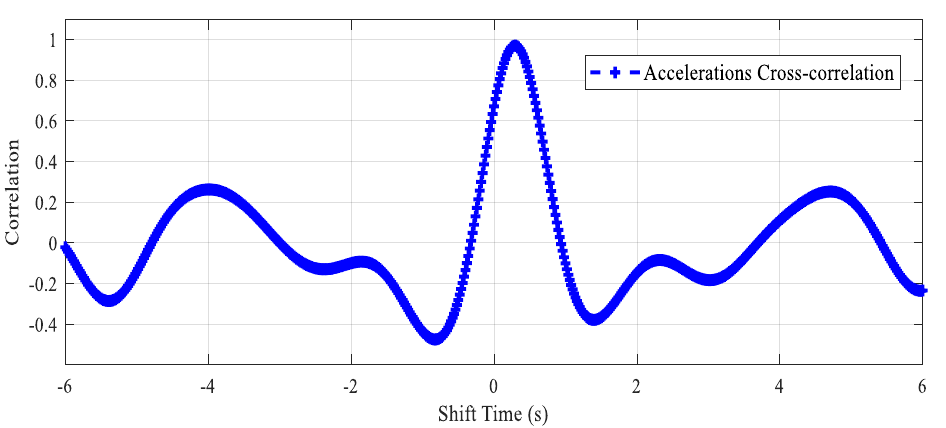}}
\caption{The cross correlation of GPS and IMU accelerations
}
\label{fig_correlation}
\end{figure}
Using augmented EKF approach needs to add the motion acceleration to the filter states to complete the relationship in the measurement function, which could make the filter unstable and introduce computational burden. Furthermore, when the GPS receiver’s update rate is relatively low, the error will be amplified in the linearization and discretization of the transformation equation. Considering all these issues, we propose another approach to estimate the time synchronization error with navigation states simultaneously, Figure \ref{fig_method} depicts the proposed method estimating time delay with another loop. After the time synchronization error$\hat{\tau }$ is estimated, three independent pure fractional delays are added to the IMU sensors’ output (gyroscope, accelerometer and magnetometer). Then, the aligned measurements are fused by the 16 states EKF, and the delayed states in $\boldsymbol x$ are estimated. To obtain the forward acceleration ${\hat{\boldsymbol a}_{fwd}}$ and the PVA states, the gyroscope and accelerometer data, which are not delayed, are used to reckon the new timely motion states. This is a discrete integral operation that applies the formula (2) (3) again, and the integral time depends on the estimated time error between IMU sensors and GPS. The dash block of figure \ref{fig_method} on the feedback loop for IMU/GPS alignment is presented in detail successively by the following three subsections.

We apply correlation of linear acceration sensed by GPS and IMU to estimate the time delay. In each direction of body coordinates, the shifting acceleration correlation between GPS and IMU is:
\begin{equation}
R(\tau )=\frac{\sum\limits_{k=0}^{n}{{{{\hat{a}}}_{fwd,k-\tau }}{{{\tilde{a}}}_{gps,k}}}}{\sqrt{\sum\limits_{k=0}^{n}{{{{\hat{a}}}_{fwd,k-\tau }}^{2}}}\sqrt{\sum\limits_{k=0}^{n}{{{{\tilde{a}}}_{gps,k}}^{2}}}}
\label{eq_crlt}\end{equation}
where the acceleration $a_{gps,k}$ is buffered by computing differential of the GPS velocity:
\begin{equation}
{{\tilde{a}}_{gps,k}}={\left( {{{\tilde{v}}}_{gps,k}}-{{{\tilde{v}}}_{gps,k-n}} \right)}/{\left( n\cdot dt \right)}
\end{equation}
where ${{\tilde{v}}_{gps,k}} $ is the newest GPS velocity, and  $dt $ is the sample period time. The delay time introduced this operation will be taken into account. The buffered GPS acceleration vector is interpolated into a new vector with equal sampling rate of IMU accelration $a_{fwd}$. The two accelerations are passed through the same smoothing filter before the cross-correlation function.

To ensure the algorithm’s convergence, before the GPS/IMU navigation system is installed on the aerial robot, an off-line calibration procedure is implemented to set a rough initial time synchronization error for on-line estimation. In this situation, the accelerations of IMU is estimated by fusing the IMU sensors (gyroscopes and accelerometers), without GPS. We make a calibration platform move in vertical when the horizontal degrees of freedom are restricted.
Figure \ref{fig_lag} shows velocities in the downward direction of North-East-Down (NED) navigation frame from the GPS receiver and the IMU. The IMU velocity is calculated by motion acceleration integral (a dead zone limitation is added to the integral calculation). It can be seen that the IMU speed error is growing with time due to the accumulation of acceleration error, and the GPS output has significant delay compared with IMU output.
 Figure \ref{fig_correlation} shows the acceleration cross-correlation shifting from -6 s to +6 s  of this off-line calibration, where the shifting time of its peak point is set as the initial rough alignment value for the time synchronization algorithm.

\subsubsection{Gain Fusion of 3-direction correlation}

Equation (\ref{eq_crlt}) can be applied to 3 directions of navigation coordinates, and by observing peak point of $R(\tau)$, 3 time delay measurements ${{z}_{\tau i}}$($i\in [1,3]$) are obtained. We use a gain fusion algorithm (GFA) \cite{b_gfa} to fuse ${{z}_{\tau i}}$, which has been used to fuse redundant sensor measurements.

\begin{table}[!ht]
\caption{GFA for fusion of redundant time delay observation  }
\label{tab_gfa}
\centering
\setlength\arrayrulewidth{0.6pt}
\begin{tabular}{lc}
\toprule[1pt]
${{{\bar{\tau }}}_{f}}={{F}_{f}}\cdot {{{\hat{\tau }}}_{f}} $ & \multicolumn{1}{c}{\multirow{3}{*}{local reset}} \\
${{{\bar{P}}}_{f}}={{F}_{f}}\cdot {{{\hat{P}}}_{f}}\cdot {{F}_{f}}^{T}+G\cdot {{Q}_{f}}\cdot {{G}^{T}}$ & \multicolumn{1}{c}{}                   \\
${{{\bar{\tau }}}_{1}}={{{\bar{\tau }}}_{2}}={{{\bar{\tau }}}_{3}}={{{\bar{\tau }}}_{f}}$ & \multicolumn{1}{c}{}                   \\ \hline
$ {{K}_{i}}=\gamma {{{\bar{P}}}_{f}}{{H}_{f}}^{T}({{H}_{f}}{{{\bar{P}}}_{f}}{{H}_{f}}^{T}+\gamma {{R}_{i}}) $ & \multirow{2}{*}{measurement update}                     \\
${{{\hat{\tau }}}_{i}}={{{\bar{\tau }}}_{f}}+V_{i}\cdot {{K}_{i}}({{z}_{\tau i}}-{{H}_{f}}{{{\bar{\tau }}}_{f}})$ &                                        \\ \hline
${{{\hat{\tau }}}_{f}}={{{\hat{\tau }}}_{1}}+{{{\hat{\tau }}}_{2}}+{{{\hat{\tau }}}_{3}}-(m-1){{{\bar{\tau }}}_{f}}$ & \multirow{3}{*}{global fusion}                     \\
${{I}_{f}}=I-{{K}_{1}}{{H}_{1}}-{{K}_{2}}{{H}_{2}}-{{K}_{3}}{{H}_{3}}$ &                                        \\
$\begin{aligned}
{{{\hat{P}}}_{f}} = & {{I}_{f}}{{{\bar{P}}}_{f}}I_{f}^{T}+{{K}_{1}}{{R}_{1}}K_{1}^{T}+\\
& {{K}_{2}}{{R}_{2}}K_{2}^{T}+{{K}_{3}}{{R}_{3}}K_{3}^{T} 
\end{aligned}$
&                                        \\ \toprule[1pt]
\end{tabular}
\end{table}

As shown in Table \ref{tab_gfa}, one iteration of time estimation loop can be divided into three main steps. The first step is to reset the local time prediction ${{\bar{\tau }}_{i}}$($i\in [1,3]$) and the prediction of covariance ${{\bar{P}}_{f}}$. Since the ${{F}_{f}}$ and $G$ are one-dimensional identity matrixes here, the prediction of GFA will degenerate into ${{\bar{\tau }}_{f}}={{\hat{\tau }}_{f}}$, and ${{\bar{P}}_{f}}={{\hat{P}}_{f}}+{{Q}_{f}}$. ${{Q}_{f}}\text{=}2.5\times 10_{{}}^{\text{-}3}$ is the process noise covariance.

In the second step, local gain ${{K}_{i}}$($i\in [1,3]$) is updated, and the delay time observation values of the GFA ${{z}_{\tau i}}$($i\in [1,3]$) are updated by correlation calculation. ${{R}_{i}}\text{=}2.25$ is the observation noise covariance. For the time estimation, ${{H}_{f}}$ is also a one-dimensional identity matrix, and the relation between state and observation is ${{z}_{\tau i}}={{\tau }_{i}}+{{e}_{\tau }}$, where ${{e}_{\tau }}$ is the observation error. Specifically, in the implementation of this algorithm, the correlation buffer length is set to 5 seconds, and the shifting window is set to 1 second to ensure the peak value is distinct in cross-correlation. For a GFA, a binary parameter $V_i$ is used to isolate the sensor fault, here the validity of the observation (cross-correlation operation) is evaluated by the MSE of the acceleration arrays and the peak value of the correlation function $R(\tau )$. Because if there is no acceleration (although this hardly occurs in each direction of a dynamic flying machine), the shifting of the cross-correlation function’s peak point will not represent the delay time. By doing so, only when both of the MSE and the peak value of the correlation function is significant, $V_i$ is assigned to 1, otherwise, it will be assigned to 0. $\gamma $ is the gain weight assigned for different sensors in GFA, since the dynamic characters in 3 directions are considered as the same, so $\gamma $ is 1/3 here.

In the third step of GFA, the covariance prediction will be corrected by three local gains ${{K}_{i}}$. The output of the GFA, ${{\hat{\tau }}_{f}}$ is the time synchronization error between the GPS data and accelerometer data from IMU.
\subsubsection{Time adjustment of IMU with delay feed back}

The last process of the time synchronization is to align the IMU data with the GPS data. An practical way is to add pure delay filters with time shifts to the IMU measurements. We re-sample the buffered IMU data with Lagrange interpolation, and then take the past value in the re-sampled buffer for EKF fusion, to realize a fractional time synchronization.

Here it is assumed that the time synchronization errors between the sensors in IMU are negligible. Therefore three equal delays could be directly add to IMU sensors’ output. However the accelerometer and gyroscope are affected by the vibration of the motors, which introduces significant high-frequency components into the frequency spectrum of the sensors’ noise, which does not satisfy the white noise assumption of Kalman filter's measurements \cite{b_wt}. So it’s essential to apply noise whitening process for the accelerometer and gyroscope based on their different noise spectral characteristics, restraining the vibration caused noise amplitude to the level of the sensors’ inherent white noise amplitude \cite{ b_accgyro}. We use finite impulse response (FIR) filters for the IMU noise whitening, and the delay introduced by these FIR filters will be compensated by:
\begin{equation}
\hat{\boldsymbol \tau }=\left[ \begin{matrix}
   {{{\hat{\tau }}}_{f}}  \\
   {{{\hat{\tau }}}_{\omega }}  \\
   {{{\hat{\tau }}}_{m}}  \\
\end{matrix} \right]=\left[ \begin{matrix}
   {{{\hat{\tau }}}_{f}}  \\
   {{{\hat{\tau }}}_{f}}+{{D}_{f}}-{{D}_{\omega }}  \\
   {{{\hat{\tau }}}_{f}}+{{D}_{f}}  \\
\end{matrix} \right]
\end{equation}
where $\hat{\boldsymbol \tau }$ is the time synchronization vector which will be used to delay the IMU data. ${{D}_{f}}$ and ${{D}_{\omega }}$ are the time delays introduced by the noise whitening filters for the accelerometer and gyroscope. ${{\hat{\tau }}_{f}}$, ${{\hat{\tau }}_{\omega }}$ and ${{\hat{\tau }}_{m}}$ are delay values on the accelerometer, gyroscope and magnet data respectively. 
\section{Experiment}
\subsection{Dynamic simulation}
\begin{figure}[!htb]
\centerline{\includegraphics[width=1.0\columnwidth]{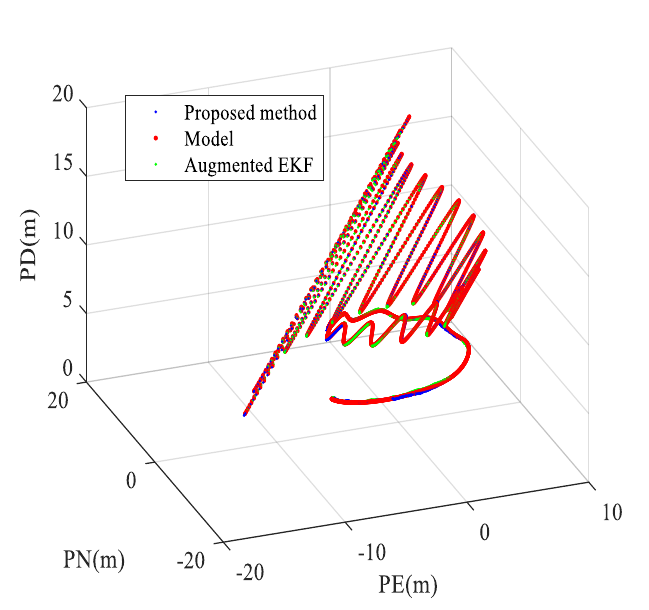}}
\caption{. 3D trajectory in 
dynamic simulation.
}
\label{fig_3D}
\end{figure}
\begin{figure*}[!htb]
\centerline{\includegraphics[width=0.8\textwidth]{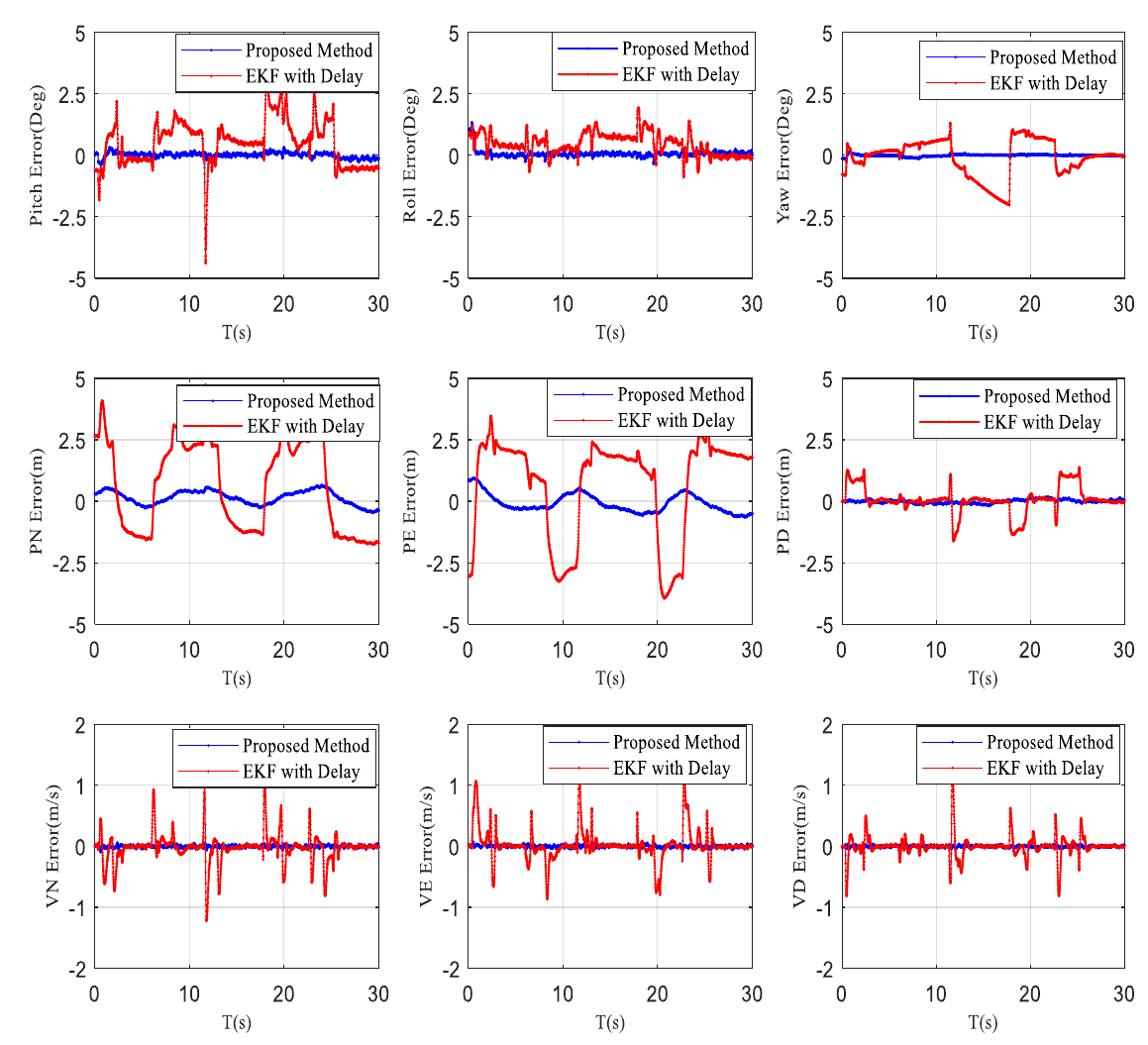}}
\caption{6-DOF motion estimation  Comparison between the proposed fusion algorithm
and the EKF without time sychronization 
}
\label{fig_9err}
\end{figure*}

\begin{table}[!htb]
\caption{Parameters settings of the simulation data generator}
\label{tab_simu}
\centering

\setlength\arrayrulewidth{0.6pt}
\begin{tabular}{lll}
\toprule[1pt]
Parameter & Value & Unit \\ \hline
gyro white noise variance      &  0.0017     & $(rad/s)^2$ \\
accelerometer white noise variance       &   0.01    &$g^2$  \\
magnetometer white unit noise variance          &  0.02     & 1 \\
max absolute gyro bias          &  0.01     &  $rad/s$\\
max absolute accelerometer bias          &  0.1     & $g$  \\
 GPS position white noise variance        &  4     & $m^2$ \\
  GPS speed white noise variance       &   0.1    &$(m/s)^2$  \\
  Max time synchronization error        &  0.2     & $s$ \\ \toprule[1pt]
\end{tabular}
\end{table}
We generate simulation data  by the following steps: first, the model trajectory and the attitude are generated within reasonable limits under a dramatically dynamic moving situation. Then based on the specifications and the noise characteristics of a low-cost MPU9250 MEMS IMU ( Invensense, 2018 ) and an off-the-shelf GPS receiver (UBLOX-NEO-M8), five groups of 3-axis sensors data are generated by the dynamic equation of the multi-rotor aerial robot. These include the angular rates, specific forces and magnetic forces in the body coordinate frame, which are measured by IMU, and the position and the linear velocity data in the navigation coordinate frame, which are outputted by GPS. The main settings on sensors output data during simulation are shown in Table \ref{tab_simu}. Figure \ref{fig_3D} shows the 3D trajectory in this simulation, where dynamic motion can be seen. The red, blue and green curves represent the true trajectory, estimated trajectory by proposed algorithm and estimated trajectory by augmented EKF respectively.
We first compare results of the proposed fusion algorithm with a common EKF fusion algorithm withoutdelay compensation. Figure \ref{fig_9err} shows the estimation errors in a period after the filter converges. It can be seen that the direct use of the EKF (red lines) has significant estimation errors in this dynamic simulation. On the contrary, the 6-DOF motion states estimated by the proposed fusion algorithm (blue lines) have higher accuracy. Furthermore, as for the proposed method, the root mean squared error(RMSE) value of estimated attitude is 0.11 degree, the RMSE value of estimated position is 0.35 $m$ and the RMSE value of estimated velocity is 0.02 ${m}/{s}$. The bias errors estimation results for inertial sensors are shown in Table \ref{tab_bias}, similarly the proposed method has high accuracy compare with EKF which does not compensate the delay. 

\begin{table}[!htb]
\caption{RMSE of the sensor bias errors estimation }
\label{tab_bias}
\centering
\setlength\arrayrulewidth{0.6pt}
\begin{tabular}{lccc}
\toprule[1pt]
Bias variables & EKF& Proposed& Unit \\ \hline
Gyroscope bias x-axis    & 0.000945 &  0.000112    & $(rad/s)$ \\
Gyroscope bias y-axis    &0.000675 &  0.000075  & $(rad/s)$  \\
Gyroscope bias z-axis    &0.000130  & 0.000080     & $(rad/s)$ \\
Accelerometer bias x-axis &  0.055240      &  0.009653     &  $g$  \\
Accelerometer bias y-axis  & 0.057460      &  0.006093    & $g$  \\
Accelerometer bias z-axis &  0.043000      &  0.005345     & $g$  \\ \toprule[1pt]
\end{tabular}
\end{table}
\begin{figure}[!htb]
\centerline{\includegraphics[width=1.0\columnwidth]{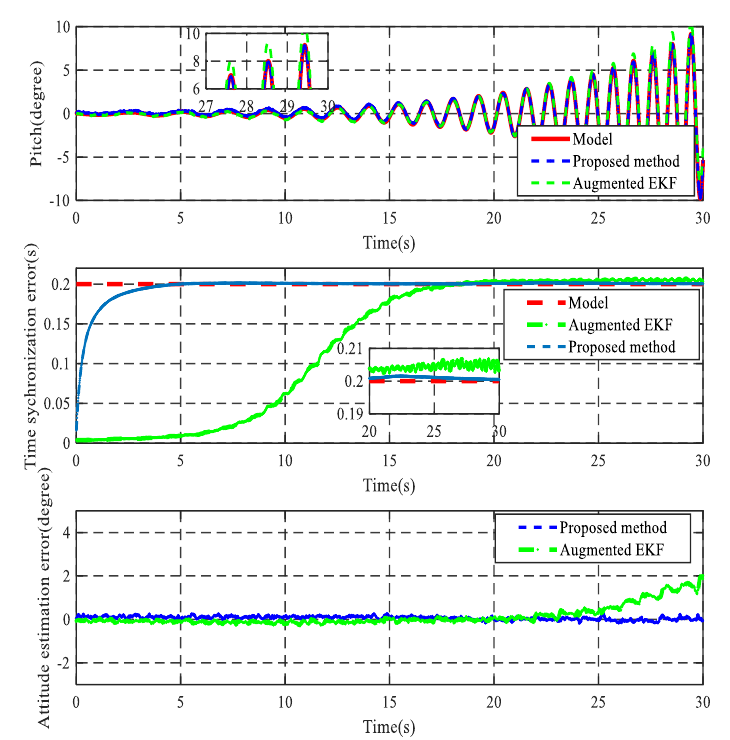}}
\caption{Comparisons of the proposed method and the augmented EKF under high dynamic situation
}
\label{fig_pva}
\end{figure}

\begin{table}[!htb]
\caption{RMSE of the estimation errors for the two methods}
\label{tab_emse}
\centering
\begin{tabular}{ccccccc}
\toprule[1pt]
 & \multicolumn{2}{c}{\begin{tabular}[c]{@{}c@{}}acceleration: \\ 19.61${m}/{{{s}^{2}}}$\end{tabular}} &  & \multicolumn{2}{c}{\begin{tabular}[c]{@{}c@{}}acceleration: \\ 50.45${m}/{{{s}^{2}}}$\end{tabular}
 } & \multirow{2}{*}{Unit} \\ \cmidrule{2-3} \cmidrule{5-6}
 & Proposed & AEKF &  & Proposed & AEKF &  \\ \toprule[1pt]
Pitch  &0.059    & 0.2237  &  & 0.0878 &1.1467 & $ ^\circ $ \\
Roll  & 0.0616  &0.2868 &  &  0.1276 &  0.8195 & $ ^\circ $ \\
 Yaw &    0.181  &  0.3116 &  & 0.1644 & 1.0829& $ ^\circ $ \\
 Velocity& 0.0163 & 0.0229&  & 0.0192 &0.0425 &  $m/s$\\
 Position& 0.3810  &  0.4462  &  & 0.0344  & 0.4896  & $m$ \\
 Delay& 0.000678 & 0.00396&  & 0.000492& 0.00506  & $s$ \\ \toprule[1pt]
\end{tabular}
\end{table}

 Figure \ref{fig_pva} shows the results of attitude angle estimation and time estimation for the augmented EKF and the proposed method. The first window shows the pitch angle estimation of the two algorithms, where the frequency and amplitude of the true attitude is increasing exponentially over time. In this circumstance, the proposed method can still estimate precisely the true value. However, the augmented EKF is less stable because the discretization error of low sample rate GPS measurement is amplified by highly dynamic motion, which can be seen in the enlarged window from 27s to 30s. Next, the second window shows the estimation results of the time synchronization error, compared with augmented EKF, the proposed algorithm has a faster convergence and better stability. The third window shows a comparison of the estimation error. Corresponding to the first window, the results of proposed method surpass the augmented EKF in higher dynamic situation (from around 20s second to 30s). Table \ref{tab_emse} compares estimation accuracy of other states with two acceleration limitations ( in a lower dynamic condition with limitation of 19.61${m}/{{{s}^{2}}}$; and a higher dynamic condition with limitation of 50.45 ${m}/{{{s}^{2}}}$). It can be seen that in both two dynamic limitation the proposed method has higher accuracy than augmented EKF, and the advantage is more obvious in higher dynamic situation.

The conclusion of the simulation is that the augmented EKF’s estimation accuracy and stability decrease as the motion dynamic intensity increases. On the contrary, the dramatic variations of the attitude and velocity have an evidently smaller impact on the estimation of the proposed method. 

\subsection{Experiments implemented on self-developed navigation system}
\begin{figure}[!htb]
\centerline{\includegraphics[width=1.0\columnwidth]{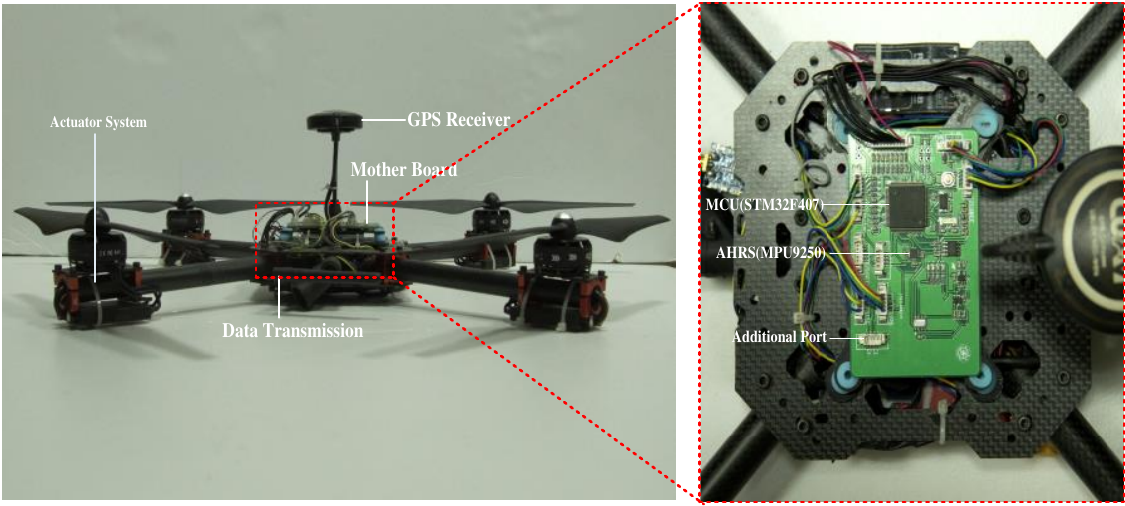}}
\caption{The UAV for experiment and the sel-designed mother board for navigation and control
}
\label{fig_uav}
\end{figure}

\begin{figure}[!htb]
\centerline{\includegraphics[width=1.0\columnwidth]{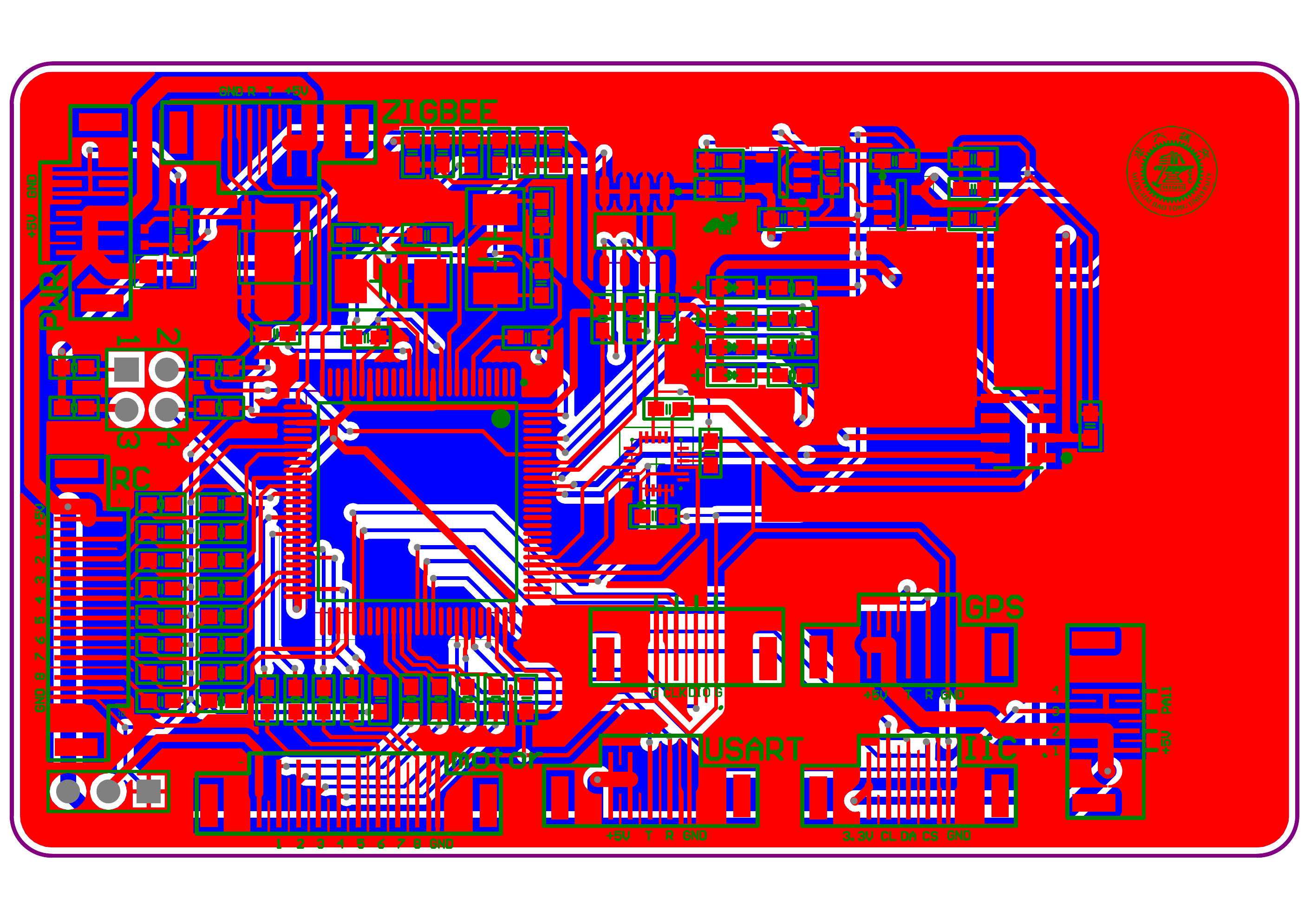}}
\caption{The designed PCB of the mother board
}
\label{fig_pcb}
\end{figure}

To validate the feasibility and effectiveness of the proposed 6-DOF estimation algorithm, the simulation code is transformed into the C++ program and realized on a microprocessor. The Microcontroller Unit (MCU) STM32f407 of the mother board is shown in Figure \ref{fig_uav}, which is leveraged by Cortex-M4 CPU with 168 MHz frequency and 128 KB RAM. The IMU MPU-9250 \cite{b_imu} is placed in the center of the Printed Circuit Board (PCB) (Figure \ref{fig_pcb}), which is constructed with MEMS 3-axis accelerometers, 3-axis gyros and 3-axis magnetometers. The resolution of these sensors in MPU-9250 is 16 bit. 

The designed mother board is set in the center of the quad-rotor aerial robot to make sure that the IMU is as close as possible to the center of rotation. The U-Blox M8M GPS/GNSS/Beidou receiver is set near to the motherboard and its horizontal and vertical distances to the center of rotation have been measured to reduce the lever-arm effect \cite{b_chang}. A Zigbee data transmission was connected to the STM32f407 to send the navigation and control data to the ground station.
The hardware system is driven by $\mu$-COS operating system. The data sample rate of MPU-9250 is set to 1000Hz, the max data update rate of U-Blox M8M GPS is 10Hz, and the data transmission rate between mother board and ground station is 100Hz. 
The processing time of propose algorithm is 8 ms per iteration, so it can realize real time 6-DOF estimation with update frequency of more than 100 Hz. 
Besides, the control algorithm is also implemented on the same microprocessor, and its accuracy requirements on the states estimation are determined by the controller design procedures (For control stability, the attitude, velocity and position estimation RMSE should be no more than 0.5 degree, 5cm/s and 1.5m respectively; the attitude and velocity sample rates should be more than 100Hz, whereas the position sample rate should be more than 50Hz to meet the flying control requirement). The estimated PVA motion states of the platform  are used directly for the system identification and flight control algorithm, and this small aerial robot was controlled to fly dynamically in an open space, where the sky was clear and there was no shelter around. 


\begin{figure}[!htb]
\centerline{\includegraphics[width=1.0\columnwidth]{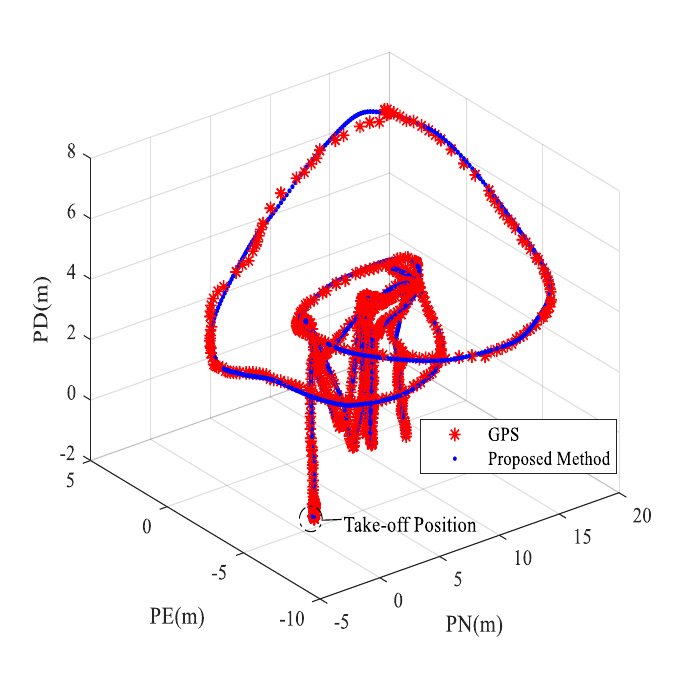}}
\caption{3D Position Estimation in flight experiment
}
\label{fig_trauav}
\end{figure}

\begin{figure}[!htb]
\centerline{\includegraphics[width=1.0\columnwidth]{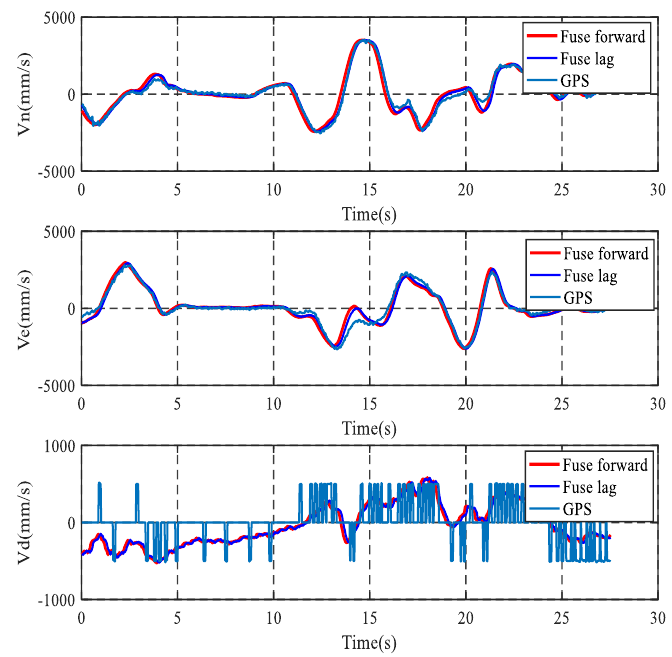}}
\caption{The two stages velocities output and the initial GPS output in flight experiment
}
\label{fig_uavatt}
\end{figure}

The result of the 3D position estimation is shown in Figure \ref{fig_trauav}. Compared with the GPS raw position trajectory, the trajectory estimated by the proposed method has higher smoothness and update rate. Figure \ref{fig_uavatt} shows the velocity estimation in a period of flight by two stages: the blue line represents the first stage which fuses the delayed IMU data and GPS data, it has a higher update rate and a higher accuracy compared to the GPS raw velocity. The red line represents the forward reckoning results based on the inertial sensors data and the output of the first stage. The second stage improves the estimation instantaneity compared to the first stage. It is worth mentioning that the GPS velocity measurement in down direction has lower accuracy than that in other two directions \cite{b_direction}, nevertheless, when the down speed relative error is significant and there are GPS measurement failures, the fusion algorithm can still be robust.

\begin{figure}[!htb]
\centerline{\includegraphics[width=1.0\columnwidth]{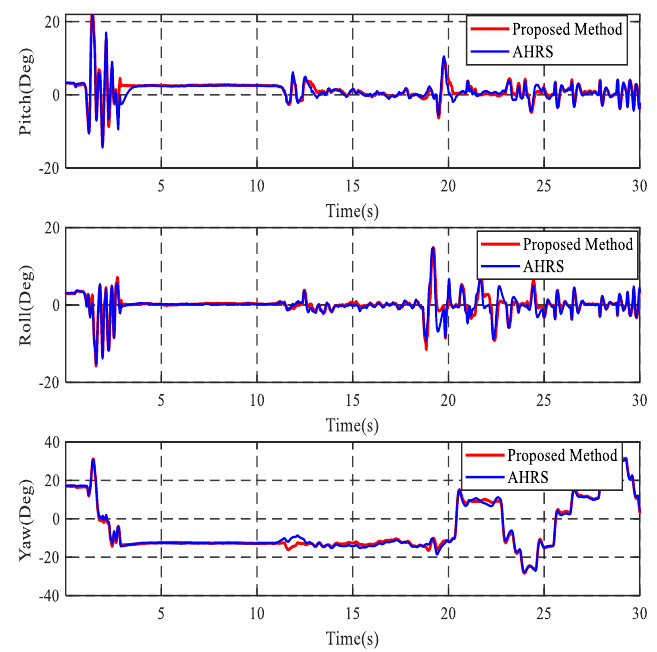}}
\caption{Attitude estimation comparisons using the proposed IMU/GPS fusion algorithm and the complementary filter based only on the IMU data.}
\label{fig_ag1}
\end{figure}

\begin{figure}[!htb]
\centerline{\includegraphics[width=1.0\columnwidth]{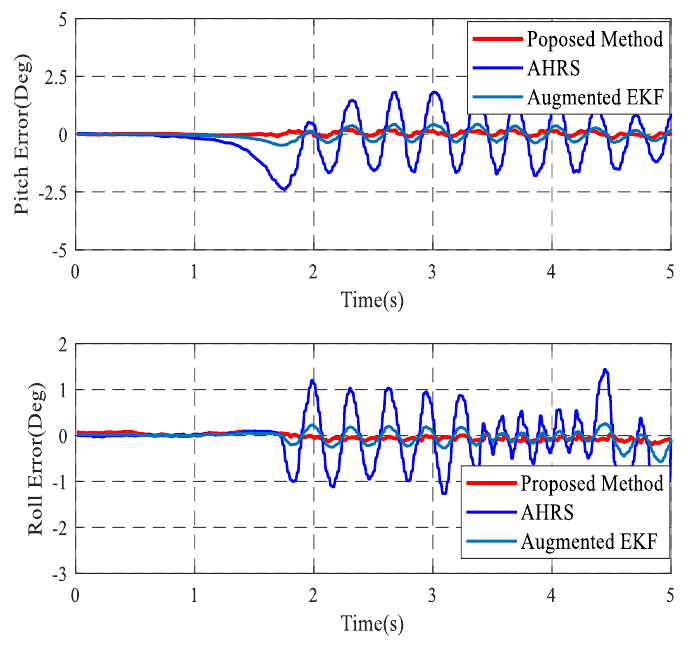}}
\caption{Attitude estimation error comparisons on dynamic validation platform}
\label{fig_ag2}
\end{figure}

For the attitude estimation, another fusion algorithm fusing gyroscope data and accelerometer data based on the concept of complementary filter \cite{b_mahony,b_comp_fusion} was employed  to fuse the IMU data without GPS data, and it runs with the IMU/GPS fusion algorithm on the microprocessor in the flight experiment. Figure \ref{fig_ag1} shows the comparisons of these two established algorithms in the flight experiment. The blue line represents the complementary filter results only using the IMU data. Although it can not remove the line accelerations’ effect on the roll and pitch estimation, it can reflect the attitude changes roughly when the aircraft linear acceleration is small. The red line shows the results of the estimated attitude using the proposed fusion algorithm, and it is followed by the blue line under the dynamic case. This indicates that this IMU/GPS fusion algorithm with time alignment has the same instantaneity and convergence as the complementary filter method based only on the IMU data.
This paper carried out a supplement experiment for estimation accuracy validation. In this experiment, the navigation system is fixed on a horizontally moving platform, and the pitch and roll angle of the navigation system is adjusted to zero by a gradienter. Figure \ref{fig_ag2} shows the pitch and roll estimation errors of three algorithms when the experimental platform is moving dynamically. Since the IMU solo method can not estimate the velocity and remove motion acceleration’s influence on attitude estimation, it has the largest estimation error (blue line), which has max estimation error of 1.5$^\circ$. The proposed method (red line) and augmented EKF ( light blue line) has the ability of linear motion estimation so that they can perform better than the IMU method. Moreover, compared with the augmented EKF which has max estimation error of 0.5$^\circ$, the proposed algorithm has higher accuracy with only max error of 0.1$^\circ$ in the dynamic situation, which also corresponds to the simulation results.

\begin{table}[!htb]
\caption{Comparison results with different delay additions}
\label{tab_delay}
\centering

\setlength\arrayrulewidth{0.6pt}
\begin{tabular}{cccccc}
\toprule[1pt]
Additional delay & $ {{\hat{\tau }}_{new}}$ & $\Delta {{\hat{\tau }}_{new}}$&$ {{\hat{\tau }}_{aekf}}$&$\Delta {{\hat{\tau }}_{aekf}}$&Unit \\ \hline
0&	0.326&	0&	0.352&	0 & $s$ \\
0.100&	0.419&	0.093&	0.438&	0.086& $s$ \\
0.200&	0.521&	0.195&	0.569&	0.215 &$s$ \\
0.300&	0.623&	0.297&	0.610&	0.258& $s$ \\ \toprule[1pt]
\end{tabular}
\end{table}
In the flight experiment, different intentional delays are added to the original GPS data. Table \ref{tab_delay} compares the estimated time synchronization errors ${{\hat{\tau }}_{new}}$ and ${{\hat{\tau }}_{aekf}}$ using the proposed algorithm and the augmented EKF respectively under different flight experiments, where different delays have been added. The differential value of the estimated delay, for the two methods, are represented by $\Delta {{\hat{\tau }}_{new}}$ and $\Delta {{\hat{\tau }}_{aekf}}$. It can be seen that compared with $\Delta {{\hat{\tau }}_{aekf}}$ of augmented EKF, the $\Delta {{\hat{\tau }}_{new}}$ is closer to the additional delay time, which indicates that the time estimation accuracy of the proposed method is higher than that of the augmented EKF in the flight experiment.
 
\section{Conclusions}
In this article, the issue of IMU/GPS integration with measurement delay is studied. The motion sensed by GPS lags behind the motion measured by IMU, and the effect of this delay on the extended Kalman filter fusion algorithm is non-negligible. Therefore this paper proposes a fusion algorithm with a time-alignment-locked loop to estimate the time synchronization error between GPS and IMU. The accuracy of the estimated states has been improved by fusing the delayed IMU data with the GPS output. Furthermore, as the estimation of the GPS delay time is based on the motion acceleration, and it’s time instantaneity is highly related to the update of measurement data, it is crucial to reckon the forward accelerations based on the estimated delayed states and the new data. This method also helps to improve the estimation instantaneity of 6-DOF motion. The results of the simulation and the flight experiments show that the time synchronization error can be well estimated by the proposed fusion algorithm. The accurate real-time estimation of the vehicle’s states ensures the disturbance rejection ability of the controller, which makes the proposed method a reliable solution for small aerial robot’s navigation. Of course, this system can also be applied for other applications such as vessels, smart car and some other platforms which need accurate 6-DOF motion information under a dynamic situation.
\section{Acknowledgments}
The present work was supported by the National Key Research and Development Project of China (201YFC0200400). The authors fully appreciate the financial support.


%


\end{document}